\documentclass[runningheads]{llncs}
\usepackage{graphicx}
\usepackage{tabu}                     
\usepackage{multirow}                 
\usepackage{multicol}                 
\usepackage{multirow}                
\usepackage{float}                    
\usepackage{makecell}                 
\usepackage{booktabs}                 
\usepackage{bm}
\usepackage{algorithm}
\usepackage{algorithmic}
\usepackage{amsmath}
\usepackage{amssymb}
\usepackage{xcolor}
\usepackage{subcaption}
\usepackage{comment}

\begin{document}
\title{D-CoRP: Differentiable Connectivity Refinement for Functional Brain Networks}
\titlerunning{D-CoRP: Connectivity Refinement for Brain Networks}
%
\author{Haoyu Hu\inst{1}\and Hongrun Zhang\inst{1}\and Chao Li\inst{1,2}\thanks{corresponding author}}
\authorrunning{H. Hu et al.}
%
\institute{Department of Clinical Neurosciences, University of Cambridge \and
Department of Applied Maths and Theoretical Physics, University of Cambridge\\\{\email{cl647@cam.ac.uk}\}}
\maketitle              
\begin{abstract}
Brain network is an important tool for understanding the brain, offering insights for scientific research and clinical diagnosis. Existing models for brain networks typically primarily focus on brain regions or overlook the complexity of brain connectivities. MRI-derived brain network data is commonly susceptible to connectivity noise, underscoring the necessity of incorporating connectivities into the modeling of brain networks. To address this gap, we introduce a differentiable module for refining brain connectivity. We develop the multivariate optimization based on information bottleneck theory to address the complexity of the brain network and filter noisy or redundant connections. Also, our method functions as a flexible plugin that is adaptable to most graph neural networks. Our extensive experimental results show that the proposed method can significantly improve the performance of various baseline models and outperform other state-of-the-art methods, indicating the effectiveness and generalizability of the proposed method in refining brain network connectivity. The code will be released for public availability.

\end{abstract}
\section{Introduction}
The human brain comprises multiple sophisticated brain regions, empowering it to navigate intricate cognitive processes and handle everyday tasks \cite{levakov2020deep}. Recent advances in neuroscience suggest that brain function is fulfilled by the coordinated activities of brain regions through connectivities \cite{park2013structural}. Brain networks, graphical representations of the brain, hold the potential to characterize regional and global brain activities \cite{bullmore2009complex,li2021brainnetgan,wei2021quantifying}. Functional brain networks, derived from function MRI (fMRI), is one of the most important tools for characterizing the brain function. However, it remains challenging to effectively leverage the information in functional brain networks due to their non-Euclidean structure.

Graph neural networks (GNNs) \cite{zhou2020graph} promise to offer effective toolkits to analyze graph data \cite{wang2021generalizable}. For example, BrainGNN is proposed to effectively aggregate the information from functional brain networks for learning tasks \cite{li2021braingnn,wei2023multi}. Graph pooling methods are developed for characterizing network modularity, e.g., TopK, a score-based pooling method  \cite{gao2019graph} and DiffPool, a hierarchical pooling method \cite{ying2018hierarchical}. However, these models only focus on node features without considering edge connections, which may particularly limit the model performance on MRI-derived brain networks that are prone to edge noises or redundancies\cite{williams2019comparison}. 

Previous studies proposed graph structure learning (GSL) to denoise and optimize graph structure  \cite{zhu2021deep}. Despite successes, existing GSL models are developed for specific type of graphs (e.g. community networks, population graphs) using rigid optimization approaches (e.g., community property optimization), and they are often limited in generalizability to other graphs once trained. Scanty models have been proposed for brain networks so far. Though a recent model \cite{zhang2023brainusl} proposes to reconstruct brain networks through time windows extracting features from fMRI, it cannot work directly on graphs (e.g., correlation matrices).

Information bottleneck (IB) theory \cite{tishby2000information} employs a probabilistic optimization between predictive performance and information compression by focusing on model-processed features of the input graph. Unlike typical GSL methods, IB is not restricted to a specific type of graph, which brings the advantages of generalizability while effectively removing noisy and irrelevant information. These advantages could equip IB as a competitive framework for denoising graphs. 

To denoise the input graph, previous IB-based methods perform subgraph filtering that updates the graph structure, e.g., subgraph information bottleneck (SIB) \cite{yu2020graph}. However, this approach treats the entire subgraph as processed features without performing differentiable refinement, which may lead to less desirable performance in removing edges. By contrast, other methods perform edge and node filtering, preserving the overall structure, e.g., VIB-GSL \cite{sun2022graph}. Nevertheless, these methods assume the optimized graph follows a standard Gaussian distribution, which may over-simplify the true distributions of edge connections that vary across different brain regions. Finally, all these methods are implemented as a fixed framework, lacking adaptability to existing GNN frameworks and hindering wider applicability. Thus, it remains a challenge to develop effective models to refine edge connections integrated into existing frameworks. 



We propose an edge-denoising method for functional brain networks based on IB theory, namely \textbf{D}ifferentiable \textbf{Co}nnection \textbf{R}efinement \textbf{P}lugin (\textbf{D-CoRP}), serving to detect and remove noisy or irrelevant connections from functional brain networks as an adaptable plugin module. Our main contributions include:

1) We design a new differentiable sampling method that renders the plugin learnable, considering the probability of completely removing noisy connections.

2) We propose a new optimization strategy to model each connection independently with multivariate distribution targets in an IB-adopted framework.


3) As far as we know, this is the first plugin for connectivity refinement of brain networks, adding minimal computation to backbone training. We also provide an efficient mode for flexible adoption. 

We evaluated our plugin on three datasets of functional brain networks. D-CoRP is shown to effectively remove noisy and irrelevant connections and enhance the performance of the GNN backbones, achieving a better and more stable performance compared with other state-of-the-art methods.

\section{Methods}
\subsection{Overview}
Define the target graph to be refined as $\bm{G}_t$, which can be the processed graph output from any layer of a graph model. Suppose $\bf{G}_t=(\bm{X}, \bm{A})$, where $\bm{X}\in \mathbb{R}^{N\times F}$ is the matrix stacking node features and $\bm{A}\in \mathbb{R}^{N\times N}$ is the adjacency matrix, with $N$ being the node number, $F$ being the dimension of node features. We aim to find a refined functional brain network adjacency matrix $\bm{A}_{re}$ based on the original $\bm{A}$, formulated as, 

\begin{equation} \label{eq_mask}
    \bm{A}_{re}=\bm{m}\odot \bm{A},
\end{equation} 

\noindent where $\bm{m} \in \mathbb{R}^{N\times N}$ is the learnable edge mask with binary elements, indicating the essential connections within brain networks. To learn the refinement mask $\bm{m}$, we propose to apply the IB theory \cite{tishby2000information,tishby2015deep}, which makes $\bm{m}$ as a processed feature connected to the model’s input and output.

\begin{figure}
\includegraphics[width=0.9\textwidth]{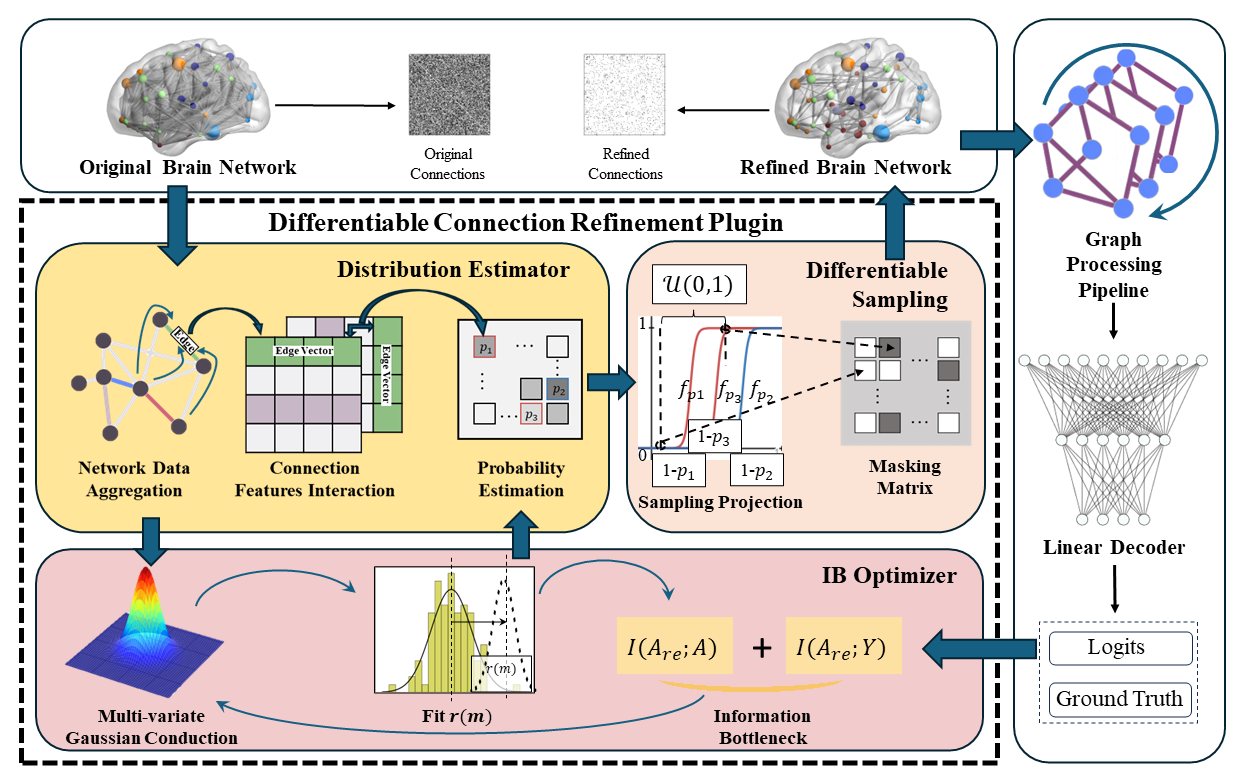}
\centering
\caption{\textbf{Model architecture.} D-CoRP refines the graph connection in two steps: 1) Differentiable sampling for the masking matrix (contains  \textbf{Distribution Estimator} and \textbf{Differentiable Sampling}); 2) Information bottleneck optimization for the refined results as shown in \textbf{IB Optimizer} part. D-CoRP can be directly plugged into the graph model as a plugin. } \label{fig1}
\end{figure}

\subsection{Differentiable Connectivity Refinement }


The binary refinement matrix $\bm{m}$ is naturally modeled as being Bernoulli distributed. However, the binary values are not differentiable, which hinders the gradient propagation in the learning process; thus, it is intractable. To address this, we relax the binary matrix $\bm{m}$ to a pseudo binary matrix, with each value in it being between $[0,1]$, which can be uniformly sampled from a predefined function. We design this function such that a sampled value will be mostly closed to 0 or 1, resembling to the binary case (see Eq.(\ref{eq:dif}) for the specific expression), which is parameterized by a learning probability matrix $\bm{\mathcal{P}}$, indicating the probabilities of corresponding edges being preserved.

\textbf{Generation of Selection Probabilities (Distribution Estimator)} The probability matrix $\bm{\mathcal{P}}$ is obtained via an auxiliary graph network (\textrm{GAT}) with the node features $\bm{X}$ and adjacency matrix $\bm{A}$, followed by the  operation $\textrm{Sigmoid}$, 
\begin{equation}
   \bm{Z}=  \textrm{GAT}(\bm{X},\bm{A}).
\end{equation}
\begin{equation}
   \bm{\mathcal{P}}=  \textrm{Sigmoid} \left( \bm{Z}\bm{Z}^T \right).
\end{equation}

\noindent where $\bm{Z}, \bm{\mathcal{P}}\in \mathbb{R}^{N\times N}$. We choose $\textrm{GAT}$ to distil critical graph information due to its efficient attention mechanism. 

\textbf{Differentiable Sampling for Masking Matrix } The sampling is performed via a differentiable function adapted from the concrete relaxation of Bernoulli distribution \cite{sun2022graph}:
\begin{equation} 
\begin{split}
    \mathcal{F}(\bm{\mathcal{P}}|\bm{A})=\mathcal{F}(\bm{\mathcal{P}},\bm{\Pi}:\mathcal{T})=\frac{1}{2}(1-\textrm{tanh}(\frac{\bm{\Pi} +\bm{\mathcal{P}}-1}{\mathcal{T}}))=\frac{1}{1+e^{-\frac{2(\bm{\Pi} +\bm{\mathcal{P}}-1)}{\mathcal{T}}}}
\end{split}
\label{eq:dif}
\end{equation}
\noindent where $\bm{\Pi} \sim \mathcal{U}(0,1)$ provides uniformly random sampling. $\mathcal{T}$ represents the temperature for the function. This function can directly relate the probability matrix to the sampling process as Bernoulli distribution. When $\mathcal{T}\rightarrow 0$, we can treat $\mathcal{F}(\bm{\mathcal{P}},\bm{\Pi}:\mathcal{T})\approx \textrm{Ber}(\bm{\mathcal{P}})$, a Bernoulli distribution, while $\mathcal{F}(\bm{\mathcal{P}},\bm{\Pi}:\mathcal{T})$ always keeps continuous in the $[0,1]$ range. Unlike the previous concrete relaxation, Eq. (\ref{eq:dif}) achieves the closed interval [0,1] for $\bm{\mathcal{P}}$, ensuring the full or zero probability of preserving brain connections. Also, it provides a smoother change of gradient without computing the logarithm, preventing the training process from exponential explosion.


\subsection{Optimization with Information Bottleneck}


In D-CoRP, the adjacency matrix $\bm{A}$ is the input data and $\bm{A}_{re}$ is the processed feature. Combined with general variational bounds developed by Alemi et al. \cite{alemi2016deep}, we design the optimization for brain network refinement (the specific derivation is provided in the supplementary): 
\begin{equation}
\begin{split}
    \mathcal{L}_{IB}=\mathbb{E}_{\bm{A}_{re}\sim p_\theta (\bm{A}_{re}|\bm{A})}[-\textrm{log}q(\bm{Y}|\bm{A}_{re})]    +\beta \textrm{KL}[p(\bm{A}_{re}|\bm{A}),r(\bm{A}_{re})]
\end{split}
\label{eq:IB}
\end{equation}

Further introducing the upper bound of Eq. (\ref{eq:IB})'s right side, we conduct two tractable optimization targets for brain network connectivity refinement:
\begin{equation}
\begin{split}
    \mathcal{L}_{AIB}=\mathbb{E}_{\bm{A}_{re}\sim p_\theta (\bm{A}_{re}|\bm{A})}[-\textrm{log}q(\bm{Y}|\bm{A}_{re})]+\beta \textrm{KL}[\mathcal{N}(\bm{\mathcal{P}}; \bm{\Sigma}),\mathcal{N}(\bm{\mu}_{\bm{m}}; \bm{\Sigma})]
\end{split}
\label{eq:aib-s}
\end{equation}
\begin{equation}
\begin{split}
    \mathcal{L}_{AIB-E}=\mathbb{E}_{\bm{A}_{re}\sim p_\theta (\bm{A}_{re}|\bm{A})}[-\textrm{log}q(\bm{Y}|\bm{A}_{re})]+\beta \textrm{KL}[p(\bm{m}|\bm{A}),r(\bm{m})]
\end{split}
\label{eq:aib-e}
\end{equation}
where $r(\bm{m})$ is the prior distribution of $\bm{m}$ approximating $p(\bm{m})$; $\bm{\mu}_{\bm{m}}$ and $\bm{\Sigma}$ are the parameters approximating the multivariate Gaussian distribution of $\bm{m}$. $\mathcal{L}_{AIB}$ is the optimization target of D-CoRP and $\mathcal{L}_{AIB-E}$ is the optimization target of D-CoRP: Efficient, a lighter version of D-CoRP. The proof can be found in the Supplementary. Since the left-sides of Eq. (\ref{eq:aib-s}) and (\ref{eq:aib-e}) are already done by the model comparing the prediction with ground truth, we only optimize an additional $\textrm{KL}[\mathcal{N}(\bm{\mathcal{P}}; \bm{\Sigma}),\mathcal{N}(\bm{\mu}_{\bm{m}}; \bm{\Sigma})]$ or $\textrm{KL}[p(\bm{m}|\bm{A}),r(\bm{m})]$. For D-CoRP: Efficient, the plugin adds an additional $\mathcal{O}(NF^2)$ in computational complexity to the plugged model.

\begin{table*}
  \centering
  \resizebox{\textwidth}{30mm}{
  \begin{tabular}{@{}ccccccccc}
    \toprule
    &&&\multicolumn{2}{c}{FCP}& \multicolumn{2}{c}{NKI-Rockland}& \multicolumn{2}{c}{ADHD-Normal}\\
    \cmidrule(r){4-5}\cmidrule(r){6-7}\cmidrule(r){8-9}
     &Baseline&Plugin &MAE$\downarrow$& RMSE$\downarrow$& MAE$\downarrow$& RMSE$\downarrow$& MAE$\downarrow$& RMSE$\downarrow$\\
    \midrule
    \multirow{12}{*}{\makecell[c]{Basic\\GNN\\Module}}&\multirow{3}{*}{GCN \cite{kipf2016semi}} &None&15.27$\pm$1.47 &21.20$\pm$6.76 &85.98$\pm$45.23 &209.49$\pm$127.55 &10.57$\pm$3.56 &18.23$\pm$5.51\\
 & &D-CoRP (Single)&\underline{8.94$\pm$0.15} &\underline{12.13$\pm$0.28}  &\underline{16.14$\pm$0.36} &\underline{20.32$\pm$1.06}  &\underline{2.77$\pm$0.23} &\underline{3.45$\pm$0.22}\\
 & &D-CoRP (Full)&9.36$\pm$0.59 &12.53$\pm$0.61 &16.76$\pm$1.22 &21.28$\pm$1.22  &2.96$\pm$0.17 &3.87$\pm$0.24 \\
 \cmidrule(r){2-9}
 & \multirow{3}{*}{GAT \cite{velivckovic2017graph}}&None &20.44$\pm$7.03 &22.08$\pm$6.87 &23.36$\pm$12.65 &28.78$\pm$12.44 &3.17$\pm$0.18 &3.59$\pm$0.24 \\
 & & D-CoRP (Single)&\underline{9.31$\pm$0.16} &\underline{13.03$\pm$0.26} &\underline{16.06$\pm$1.50} &\underline{20.18$\pm$1.51}  &\underline{2.83$\pm$0.19} &\underline{3.47$\pm$0.23} \\
 & & D-CoRP (Full)&9.80$\pm$0.74&13.37$\pm$1.08  &18.00$\pm$1.45 &22.26$\pm$2.70 &3.28$\pm$0.27 &4.20$\pm$0.30 \\
  \cmidrule(r){2-9}
 & \multirow{3}{*}{GIN \cite{wu2020graph}}&None &10.85$\pm$0.38 &14.76$\pm$1.71  &17.43$\pm$4.12 &20.28$\pm$3.68 &5.92$\pm$1.47 &8.90$\pm$3.38 \\
 & &D-CoRP (Single) &\underline{9.11$\pm$0.18} &\underline{12.24$\pm$0.38}  &15.62$\pm$1.43 &19.60$\pm$1.66  &\underline{2.73$\pm$0.13} &\underline{3.42$\pm$0.17} \\
 & &D-CoRP (Full) &9.29$\pm$0.29 &12.50$\pm$0.24 &\textbf{15.31$\pm$0.73} &\textbf{19.06$\pm$0.97}  &3.18$\pm$0.41 &3.89$\pm$0.46 \\
 \cmidrule(r){2-9}
&\multirow{3}{*}{GraphSAGE \cite{hamilton2017inductive}} &None &15.72$\pm$2.51 &30.24$\pm$11.99  &44.55$\pm$14.64 &104.86$\pm$60.86  &4.42$\pm$0.26 &5.44$\pm$0.32 \\
 & &D-CoRP (Single) &\underline{9.28$\pm$0.39} &\underline{12.33$\pm$0.47}  &\underline{15.19$\pm$0.91} &\underline{19.71$\pm$1.42}  &\textbf{2.65$\pm$0.17} &\textbf{3.21$\pm$0.18} \\
 & &D-CoRP (Full) &9.35$\pm$0.50 &12.70$\pm$0.50  &17.06$\pm$1.77 &20.71$\pm$2.01  &2.87$\pm$0.25 &3.57$\pm$0.30 \\
 \hline
\multirow{6}{*}{\makecell[c]{Basic\\Graph \\ Pooling}}& \multirow{3}{*}{DiffPool \cite{ying2018hierarchical}}&None &12.10$\pm$0.79 &18.01$\pm$4.55 &26.96$\pm$9.63 &39.04$\pm$19.44  &4.06$\pm$0.21 &6.56$\pm$0.86 \\
& &D-CoRP (Single)&\textbf{8.55$\pm$0.17} &\textbf{11.83$\pm$0.60}  &\underline{16.32$\pm$2.44} &\underline{20.39$\pm$2.32}  &\underline{2.99$\pm$0.20} &\underline{3.75$\pm$0.18} \\
 & &D-CoRP (Full) &8.84$\pm$0.70 &11.96$\pm$0.73  &24.62$\pm$10.24 &28.56$\pm$10.85  &3.08$\pm$0.26&3.89$\pm$0.36 \\
 \cmidrule(r){2-9}
&\multirow{3}{*}{TopK \cite{gao2019graph}} &None &\underline{11.05$\pm$0.69} &\underline{13.23$\pm$1.11}  &27.49$\pm$9.42 &37.61$\pm$10.82  &6.14$\pm$3.62 &6.70$\pm$3.42 \\
 & &D-CoRP (Single) &12.22$\pm$6.40 &15.21$\pm$5.72 &\underline{19.88$\pm$8.72} &\underline{24.10$\pm$9.11}  &\underline{2.73$\pm$0.20} &\underline{3.78$\pm$0.10} \\
 & &D-CoRP (Full) &13.98$\pm$4.46 &16.34$\pm$4.48  &30.13$\pm$8.31 &35.16$\pm$9.36  &2.84$\pm$0.20 &3.58$\pm$0.28 \\
    \bottomrule
  \end{tabular}}
  \caption{\textbf{Performance of D-CoRP on  Baselines}. Underlines indicate local optimal comparing None  (D-CoRP), D-CoRP (Single) (added to a single GNN block), and D-CoRP (Full) (added to all GNN blocks). Underlines indicate optimal results of corresponding baselines and bold indicates global optimal results.}
  \label{tab:example}
\end{table*}

\section{Experiments and Results}
\subsection{Task and Datasets}
We focus on brain age prediction as the primary task. Brain age is shown as associated with brain network features \cite{stanley2015changes,bashyam2020mri}, serving as an ideal benchmark for model evaluation. The resting-state fMRI data from three datasets were used: 1) 1000 Functional Connectomes Project (FCP)  \cite{biswal2010toward} (1,003 healthy subjects aged 18-85 yrs (mean 28$\pm$13 yrs, 569 females). 2) Enhanced Nathan Kline Institute - Rockland Sample (NKI-Rockland) (393 subjects, mean 35$\pm$20 yrs, 164 females) \cite{nooner2012nki,tobe2022longitudinal}, known for wide age range (4-85 yrs). 3)  Normal subjects (no diseases) from Attention-Deficit Hyperactivity Disorder-200 (ADHD-Normal)  (330 subjects, mean 12$\pm$3 yrs, 166 females) \cite{adhd2012adhd}, focusing on the younger group.   All datasets are available in the UCLA multimodal connectivity database \cite{brown2012ucla}.

\begin{figure*}
  \centering
  \includegraphics[width=\linewidth]{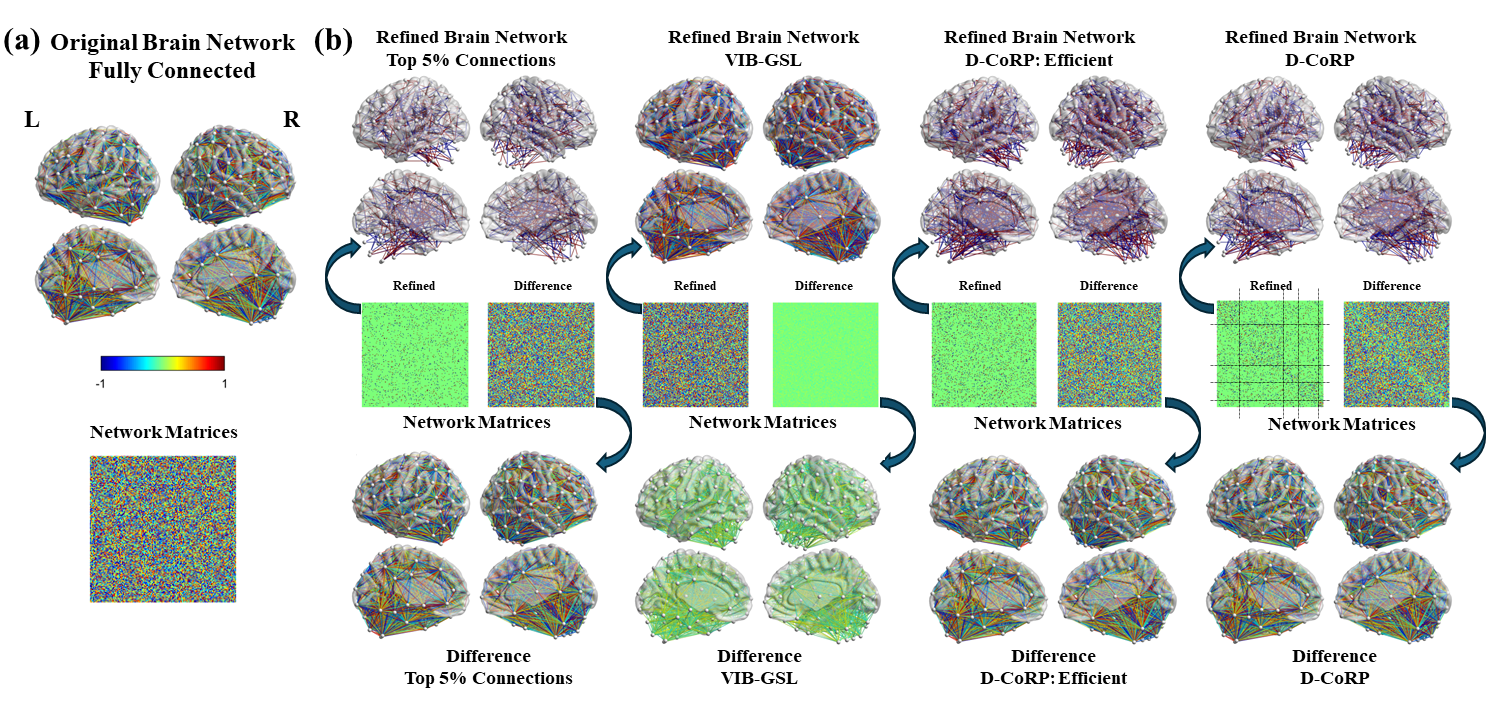}
  \caption{\textbf{Visualization of Refined Brain Network.} One example of the brain networks refined by D-CoRP, compared to other denoising methods all based on GCN. (a): original brain network; (b): refined brain networks and their differences from the original network with various connectivity refinement methods.}
  \label{fig:brain_2}
\end{figure*}
\subsection{Experiment Settings and Evaluation Metrics}

We use two GNN blocks and a multi-layer perceptron (MLP) as the backbone (see Supplementary for detailed structures). We used pooling methods (TopK and DiffPool) with a pooling ratio of 0.25.

In section 3.3, to test the adaptability of D-CoRP, we plug it into 1) the initial part of the model (D-CoRP (Single)), which refines the original brain network; 2) every GNN block of the model (D-CoRP (Full)), which refines the original and processed brain network. The $\bm{\Sigma}$ is set as $\bm{I}_{N^2\times N^2}$, $\bm{\mu}_{\bm{m}}$ is set as $\frac{\bm{1}_{N^2}}{20}$, where $\bm{1}_{N^2}$ is $N\times N$ matrix with value 1 in each place; $\mathcal{T}=0.01$, and $\beta=1$. In section \ref{sec:comparison}, for D-CoRP: Efficient, $r(\bm{m})$ is set as a fixed probability, 0.05. 

All training is performed on an Nvidia A-100 with pytorch 2.1 and pytorch-geometric package \cite{fey2019fast}. We choose Adam as the optimizer with a cosine annealing learning rate $\frac{1}{10}\textrm{cos}(\frac{epoch_c}{epoch_T})$, where $epoch_c$ stands for the current epoch number counting from 0 and $epoch_T$ stands for the training epoch set as 150. MSE is used as the loss function. 10-fold cross-validation is applied to all datasets.

Two performance metrics are adopted for evaluation: Mean Absolute Error (MAE) and Root Mean Square Error (RMSE), i.e., $\textrm{MAE}=\frac{\sum_{n=1}^N |y_i-\hat{y_i}|}{N}$,  $\textrm{RMSE}=\sqrt{\frac{\sum_{n=1}^N (y_i-\hat{y_i})^2}{N}}$, where $y_i$ is the ground truth and $\hat{y_i}$ is the predicted value. 


\subsection{Performance of D-CoRP on Baselines}
\label{sec:backbone}
To test the applicability of D-CoRP, we tested the most widely used GNN baselines: GCN \cite{kipf2016semi}, GAT \cite{velivckovic2017graph}, GIN \cite{wu2020graph}, and GraphSAGE \cite{hamilton2017inductive}. Due to the importance of graph pooling methods in brain network research\cite{vidaurre2017brain,hilgetag2020hierarchy}, we also tested  TopK pooling and DiffPool\cite{gao2019graph,ying2018hierarchical}, known for their capability to distil subgraphs from brain networks. 

The experimental results (Table \ref{tab:example})  show that D-CoRP significantly outperforms baseline models in all comparisons, supporting the effectiveness of D-CoRP. Of note, with D-CoRP integrated, the backbones exhibit reduced standard deviations in the assessment metrics, suggesting improved stability. This improvement is particularly relevant in brain network research due to data heterogeneity arising from demographics and acquisition. Our results also reveal that adding D-CoRP to all the GNN blocks, i.e., D-CoRP (Full), however, does not perform better in most cases, indicating over-compressed information.
\begin{table*}
  \centering
  \resizebox{\textwidth}{27mm}{
  \begin{tabular}{cc@{}ccccccc}
    \toprule
    &&&\multicolumn{2}{c}{FCP}& \multicolumn{2}{c}{NKI-Rockland}& \multicolumn{2}{c}{ADHD-Normal}\\
    \cmidrule(r){4-5}\cmidrule(r){6-7}\cmidrule(r){8-9}
     \multicolumn{2}{c}{Method}&Baseline &MAE$\uparrow$& RMSE$\downarrow$& MAE$\downarrow$& RMSE$\downarrow$& MAE$\downarrow$& RMSE$\downarrow$\\
    \midrule
&\multirow{3}{*}{N/A-5\%} &GCN&18.62$\pm$4.27 &30.22$\pm$7.36 &55.35$\pm$10.44 &82.12$\pm$24.45 &11.42$\pm$5.86 &22.21$\pm$6.52\\
 & &GAT&21.30$\pm$2.88&46.66$\pm$2.98 &44.43$\pm$7.46 &62.27$\pm$8.69 &16.32$\pm$2.42 &20.78$\pm$3.11\\
 & &GIN&21.80$\pm$2.32 &25.99$\pm$2.57 &30.69$\pm$3.46 &38.33$\pm$4.21  &9.45$\pm$2.95 &15.74$\pm$4.01 \\
 \cmidrule(r){1-9}
 & \multirow{3}{*}{SIB \cite{yu2020graph}}&GCN&27.24$\pm$0.15 &30.06$\pm$0.20 &33.96$\pm$1.72 &39.39$\pm$1.46 &11.56$\pm$0.22 &12.01$\pm$0.23 \\
 & &GAT&27.41$\pm$0.22 &30.21$\pm$0.41 &35.62$\pm$0.92 &41.18$\pm$1.23  &11.80$\pm$0.43 &12.25$\pm$0.98 \\
 & &GIN&27.28$\pm$1.25&29.81$\pm$1.31  &35.40$\pm$1.66 &40.80$\pm$2.53 &11.23$\pm$1.45 &11.67$\pm$1.64 \\
  \cmidrule(r){1-9}
 & \multirow{3}{*}{VIB-GSL \cite{sun2022graph}}&GCN&\textbf{8.66$\pm$0.40} &13.90$\pm$0.58  &16.44$\pm$1.88 &21.02$\pm$1.63 &4.66$\pm$1.68 &5.73$\pm$1.54 \\
 & &GAT&11.10$\pm$0.94 &17.10$\pm$0.85 &16.69$\pm$1.43 &22.01$\pm$1.43 &4.55$\pm$0.76 &5.55$\pm$0.75 \\
 & &GIN&12.54$\pm$0.46 &17.78$\pm$0.61 &30.58$\pm$1.71 &36.33$\pm$2.02  &8.92$\pm$0.80 &9.66$\pm$0.79 \\
 \cmidrule(r){1-9}
 &\multirow{3}{*}{\textbf{D-CoRP: Efficient}} &GCN&11.50$\pm$0.96&16.66$\pm$0.98 &34.23$\pm$1.33 &41.07$\pm$3.17 &5.57$\pm$1.36 &10.68$\pm$2.01 \\
 & &GAT&12.96$\pm$2.07 &20.41$\pm$4.65 &36.71$\pm$1.93 &41.53$\pm$2.42  &\textbf{2.53$\pm$0.41} &\textbf{2.96$\pm$0.44} \\
 & &GIN&11.76$\pm$1.08 &16.22$\pm$1.16 &16.59$\pm$0.77 &20.00$\pm$1.00 &9.66$\pm$2.01 &15.93$\pm$3.08 \\
 \cmidrule(r){1-9}
&\multirow{3}{*}{\textbf{D-CoRP}} &GCN&8.94$\pm$0.15 &\textbf{12.13$\pm$0.28}  &16.14$\pm$0.36 &20.32$\pm$1.06  &2.77$\pm$0.23 &3.45$\pm$0.22 \\
 & &GAT&9.31$\pm$0.16 &13.03$\pm$0.26 &16.06$\pm$1.50 &20.18$\pm$1.51  &2.83$\pm$0.19 &3.47$\pm$0.23 \\
 & &GIN&9.11$\pm$0.18 &12.24$\pm$0.38  &\textbf{15.62$\pm$1.43} &\textbf{19.60$\pm$1.66}  &2.73$\pm$0.13 &3.42$\pm$0.17 \\
\bottomrule
\end{tabular}}
  \caption{\textbf{Comparison of D-CoRP and Other Denoising Methods.} The bold font indicates the global optimal result for each dataset.}
  \label{tab:comparison}
\end{table*}
\subsection{Comparison with Other Denoising Methods}
\label{sec:comparison}
We compare D-CoRP and D-CoRP: Efficient with conventional denoising method, selecting top 5\% connections according to correlation values. We also compare two IB-based graph structure learning methods, i.e., SIB \cite{yu2020graph}, and VIB-GSL \cite{sun2022graph}. All comparisons involve three different backbones (i.e., GCN, GAT and GIN)  as the graph processing module.

From Table \ref{tab:comparison}, we observe that D-CoRP outperforms other models in most comparisons. In all cases, D-CoRP shows better stability over the comparison models, indicated by lower variance. These results support the effectiveness of D-CoRP for brain network research. Additionally, our results imply that D-CoRP: Efficient may be more suitable for smaller datasets (e.g., ADHD-Normal).

\begin{figure}
\centering
\begin{subfigure}[b]{0.4\textwidth}
\includegraphics[width=\textwidth]{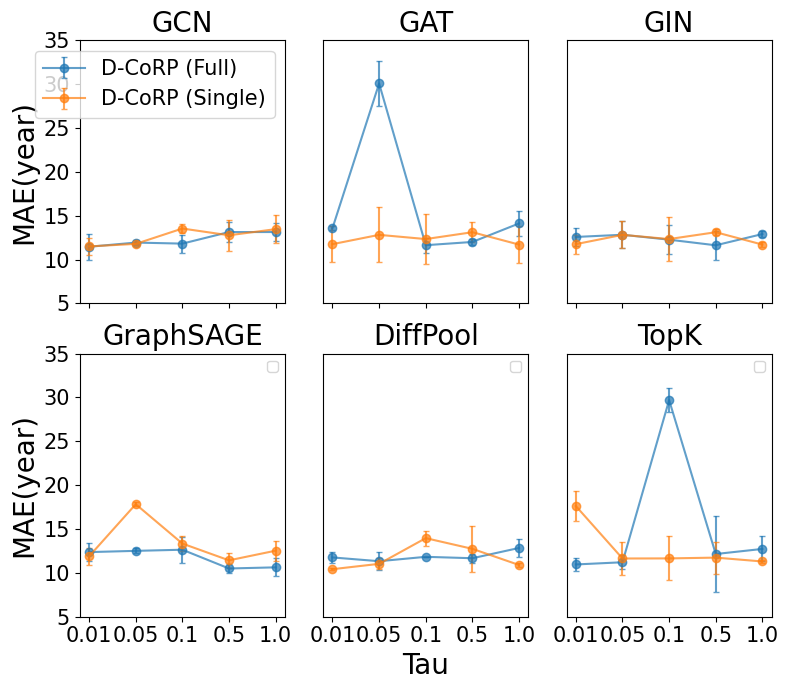}
\caption{\textbf{D-CoRP: Efficient}}
\end{subfigure}
\begin{subfigure}[b]{0.4\textwidth}
\includegraphics[width=\textwidth]{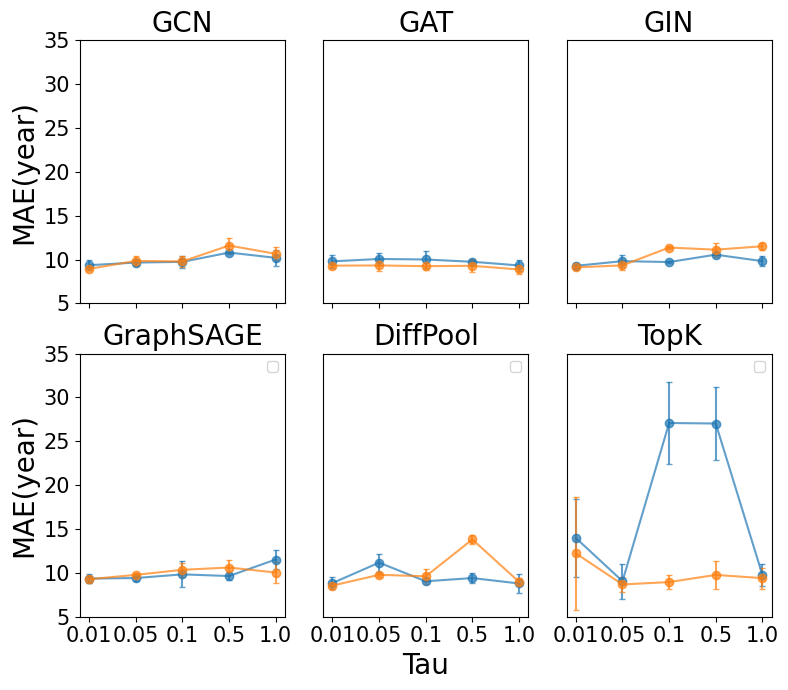}
\caption{\textbf{D-CoRP}}
\end{subfigure}
\caption{\textbf{Results of Variant Temperature.}} \label{fig4}
\end{figure}

Further, we visualize an example network derived from the ADHD-Normal dataset in Fig. (\ref{fig:brain_2}). The results show that our methods effectively remove noisy connections compared to other methods. Interestingly, from the refined network from D-CoRP, we observe the modularity preserved in the functional network, one of the most important brain network properties, suggesting the D-CoRP's capability to preserve informative edges alongside denoising.
\subsection{Hyperparameter Tuning on D-CoRP}
The temperature $\mathcal{T}$ controls the slope of sampling function $\mathcal{F}(\bm{\mathcal{P}},\bm{\Pi}:\mathcal{T})$, influencing how effectively D-CoRP refines brain network connectivities. Six backbone models in Table. \ref{tab:example} are utilized to test the effectiveness of our D-CoRP. 10-fold cross-validation is performed on the FCP dataset. MAE is used to evaluate the final performance. Fig. \ref{fig4} shows $\mathcal{T}$ value of 0.01 for D-CoRP may lead to superior and stable model performance in more than half of cases and setting $\mathcal{T}$ as 1 may be generally suitable for all cases. We recommend setting $\mathcal{T}$ as 0.01 if only binary masking is needed while setting $\mathcal{T}$ as 1 if D-CoRP is also utilized to adjust the importance of connections in addition to the denoised part.
\section{Conclusion}
In this study, we propose D-CoRP, a tool for differentiable connectivity refinement in functional brain networks based on information bottleneck theory. Specifically, we design a differentiable sampling function and multivariate probability modeling for detecting valuable connections in the brain network. Our experiments show that D-CoRP achieves state-of-the-art performance in connectivity refinement. As a plugin, D-CoRP can be integrated into existing GNN-based frameworks and provides an efficient mode for flexible adoption. Future work will develop multivariate optimization with adaptive covariance tailored to functional connectivity.






%
%
%
%





{
\bibliographystyle{splncs04}  
\bibliography{MICCAI}}

\end{document}